\documentclass[sigconf,nonacm]{acmart}
\usepackage{tikz}
\usetikzlibrary{positioning,calc}


\begin{document}

\title{Scalable Conversational Assessment: Chat-Based Exams with a Two-Agent LLM Pipeline}

%% arXiv version: real author block restored (the submission used "Author One"/"Author
%% Two" placeholders for SIGCSE's double-anonymous review).
\author{Geoffrey Challen}
\orcid{0000-0002-7249-7867}
\affiliation{%
  \institution{University of Illinois Urbana-Champaign}
  \city{Urbana}\state{Illinois}\country{USA}}
\email{challen@illinois.edu}
\author{Craig Zilles}
\affiliation{%
  \institution{University of Illinois Urbana-Champaign}
  \city{Urbana}\state{Illinois}\country{USA}}
\email{zilles@illinois.edu}
\renewcommand{\shortauthors}{Challen and Zilles}

\begin{abstract}
Oral exams are compelling because (1) students can be asked to provide their reasoning for an answer and (2) their interactive nature permits follow-up questions to resolve ambiguous answers.
But oral exams scale poorly.
In this paper, we present \emph{scalable conversational assessment}, an LLM-based system that conducts interactive text-based conversations with students to assess their ability to make and support an argument.
We use a two-agent architecture to separate concerns and reduce the likelihood that answers are leaked to the student: the \emph{interviewer} interacts directly with the student, but only receives guidance about the next question to ask the student; the hidden \emph{evaluator} knows the goals of the whole activity, tracks the student's progress using a rubric, and guides the interviewer on when to follow-up on the current question or proceed to the next question.
The system was first tested using a collection of simulated students with LLM personas of a particular ability level or adversarial tactic (e.g., answer-extractors, prompt-injectors, confident fabricators).
A novel element of this evaluation was that it required the use of a fictional assessment topic to prevent personas that shouldn't have information from using knowledge present in the LLM's pre-training.
The system was also deployed in a small, elective AI literacy course in both formative and summative contexts.
Through manual and post-facto AI review of the transcripts we identified opportunities to improve the follow-up question generation and ensure that grading only credited ideas that the student (and not the interviewer) contributed.

\end{abstract}

\maketitle

\section{Introduction}
\label{sec:intro}

Most instructors intuitively recognize the benefit of oral exams.
Talking with a student remains the most direct way to find out whether they can \emph{explain} what they know, and any ambiguities can be clarified through follow-up questions.
Instructors who use oral exams report exactly that benefit, along with strikingly positive student reactions~\cite{ohmann2019oral, ohmann2025oral, reckinger2022oral}.
In addition, oral exams conducted in person is one of the few assessment formats that an LLM cannot complete on a student's behalf~\cite{ohmann2025oral, novak2025chatgpt}.

Yet oral exams remain rare, for two stubborn reasons.
The first is cost: examiner time grows linearly with enrollment.
One recent report spent twenty hours of a final-exam week examining forty students~\cite{ohmann2025oral}, an earlier study judged regularly scaling beyond twenty or thirty students ``challenging''~\cite{ohmann2019oral}, and running oral exams in courses of hundreds has used teams of trained teaching assistants and binary pass/fail rubrics~\cite{reckinger2022oral, reckinger2021highenrollment}.
The second is consistency.
Every student inevitably receives a slightly different exam~\cite{ohmann2019oral}, and a century of evidence says human evaluation of open-ended performance is noisy: graders disagree with one another and with themselves~\cite{brookhart2016century, starch1913history, eells1930reliability, brimi2011reliability}, examiner identity alone flips outcomes in professional clinical examinations~\cite{mcmanus2006examiner}, and 28 CS1 graders sharing a rubric achieved near-zero reliability~\cite{messer2025consistent}.
Practitioners' main defense, scripting the exam so tightly that the proctor reads from a rubric~\cite{ohmann2025oral}, buys consistency by giving up much of the conversational flexibility that motivated the format.

This paper reports our experience building and deploying a \emph{scalable conversational assessment}: an assessment conducted as a real-time typed conversation with a pair of cooperating AI agents (Section~\ref{sec:design}).
The \emph{interviewer} conducts the conversation with the student with only guidance of what it should ask next, so as to avoid leaking answers to the student.
A hidden \emph{evaluator} has full access to the assignment's goals and rubrics, tracks and grades the student's progress, and guides the interviewer.
To be clear about modality, nothing is spoken; these are chat-based exams that pursue the oral exam's goals without its mechanics.

Before deploying in the classroom, we validated the system with LLM-simulated students, both legitimate and adversarial (Section~\ref{sec:validation}).
This evaluation necessitated the use of a deliberately fictional topic so that persona knowledge cannot leak in from model pre-training.
We did a test deployment of the system through multiple activities in a small enrollment AI literacy course (Section~\ref{sec:deployment}).
It was first introduced as voluntary practice in February 2026 and culminated in an end-of-semester pilot as a proctored computer-based exam in May 2026.
We did both a manual review of a subset of the transcripts and a full post-facto LLM-based review of the transcripts using a different model to evaluate both the interviewer and the evaluator's grading.
Through the use of genAI, this approach is inherently scalable, and, because the same evaluator examines every student, it avoids the inconsistency that arises from multiple examiners.
In our classroom deployment, students covered three distinct topics in a twenty-minute conversation, at a cost of about a quarter per assessment.
In Section~\ref{sec:discussion}, we discuss our findings and planned revisions for our next deployment.
In a final version of this paper, we will provide a public GitHub repository that will allow replicating our two-agent architecture along with an example assignment and its rubric.

\section{Background and Related Work}
\label{sec:background}

Three bodies of work frame this paper: the oral-exam tradition we draw our goals from, the evidence that human judgment of open-ended performance needs validating, and the emerging practice of AI graders and simulated students.

\subsection{Oral Exams in Computing Education}

Oral assessment in CS has a small but growing literature.
Gharibyan compared oral and written exams two decades ago~\cite{gharibyan2005assessing}; the pandemic prompted renewed experimentation~\cite{sabin2021oral, lubarda2021oral}; and Ohmann and Novak developed and evaluated a scripted-rubric oral final across institutions, finding positive student sentiment and instructor confidence in what the exams measure~\cite{ohmann2019oral, ohmann2025oral}.
Reckinger and Reckinger scaled TA-administered oral exams to courses of hundreds, with experience positive across demographic groups and stress falling after students' first exam~\cite{reckinger2022oral, reckinger2021highenrollment}.
The LLM era has made the integrity argument central~\cite{novak2023oral, novak2025chatgpt}.
Across this literature the same two limits recur: examiner hours scale with enrollment, and consistency depends on scripting.

\subsection{Humans Are Noisy Graders}

Most of the research measuring the lack of grading reliability come from outside of CS.
Starch and Elliott showed in 1912--1913 that teachers grading identical exams spread their marks across dozens of points, concluding the variability was ``not a function of the subject but a function of the examiner and of the method of examination''~\cite{starch1912reliability, starch1913reliability, starch1913history}; graders also disagree with themselves on re-marking~\cite{eells1930reliability}, training does not eliminate the spread~\cite{brimi2011reliability, brookhart2016century}, and the pattern holds for expert judgment generally~\cite{kahneman2021noise, dawes1989clinical, grove2000clinical}.
Examiner stringency alone (the ``hawk--dove effect'') flipped about one pass/fail outcome in twenty-five in the UK's MRCP clinical examinations~\cite{mcmanus2006examiner}, and paired examiners scoring identical performances showed low concordance across five medical schools~\cite{schleicher2017examiner}.
In computing education, Messer et al.\ found near-zero inter-rater reliability among 28 graders sharing a rubric, individually inconsistent even on re-grading their own work~\cite{messer2025consistent}.
Our aim is not to claim AI grading is flawless, but to reject the baseline that human judgment of conversational performance is reliable: any examiner, human or AI, must be \emph{validated}, not verified.

\subsection{AI Graders and Simulated Students}

LLM-assisted grading of open-ended work is advancing in CS education: LLM-as-judge evaluation of programming feedback~\cite{koutcheme2025evaluating} and rubric-anchored grading of free-response questions, where the AI also refines the rubric~\cite{zhao2026grading}.
Our system differs: the AI does not just grade an artifact but conducts the examination, deciding in real time what to ask next, while a separate agent grades.
Validating with LLM-simulated students also has precedent~\cite{dickey2026glow}; our methodological contribution is simulating them against a \emph{fictional} topic, so the knowledge level is controlled by the persona prompt rather than contaminated by pre-training (Section~\ref{sec:validation:fictional}).

\section{System Design}
\label{sec:design}

The system was first conceived as a conventional AI-assisted grading of student writing.
Efforts to shorten the feedback loop led us to question why the feedback was not simply a conversation, and the analogy to oral examination structured all that followed.

The resulting conversational assessment is a chat-based exam.
Instead of answering multiple-choice questions or writing essays, the student holds a real-time typed conversation with an AI interviewer that asks questions, follows up on interesting points, and probes for deeper understanding.
The goal is to evaluate whether students can \emph{explain} what they know.

\begin{table}
  \caption{Example questions from the deployed assessments.}
  \vspace{-3mm}
  \label{tab:questions}
  \small
  \begin{tabular}{@{}lp{0.68\columnwidth}@{}}
    \toprule
    Kind & Example \\
    \midrule
    Explain & How does the Turing test work, and what does passing mean? \\
    Explain & Name a concept about how AI works that you learned this
              semester and explain it in your own words. \\
    Defend  & Do current AI models pass the Turing test? \\
    Defend  & Can AI be genuinely creative? \\
    Reflect & What belief about AI did you hold entering the course, and
              how, if at all, has it shifted? \\
    Reflect & What practice or boundary for your own AI use will support
              your goals and well-being going forward? \\
    \bottomrule
  \end{tabular}
\end{table}

Our assessments typically consisted of multiple \emph{portions}.
We experimented with the three portion types shown in Table~\ref{tab:questions}.
\emph{Explain} portions ask the student to articulate how something works, checked against provided reference material.
\emph{Defend} portions ask the student to take a position and defend it; they are graded not on which position is taken but on the quality of reasoning and engagement with counterarguments.
Simply conceding a counterargument does not count as engagement.
\emph{Reflect} portions, which dominated our end-of-semester quiz, have no reference answer at all: instead, students explain how a concept changed their thinking, a concern they hold, or a practice they intend to adopt, and they are graded on specificity and reasoning.

\begin{figure}
  \small
  \setlength{\fboxsep}{6pt}%
  \fbox{\parbox{\dimexpr\columnwidth-2\fboxsep-2\fboxrule}{%
  \textbf{Portion} (factual): \emph{Explain the Turing test.}

  \smallskip
  \textbf{Criteria:} (1)~the setup; (2)~the purpose, testing machine
  intelligence by imitation; (3)~what passing means, indistinguishable from a
  human.

  \smallskip
  \textbf{Criterion 1, three levels:} \texttt{met} --- volunteers the
  three-participant setup (interrogator, human, machine) \emph{and} text-only
  communication, unprompted; \texttt{partially\_met} --- names some participants
  but not all three, or only after being asked; \texttt{not\_started} --- no
  setup, or a fundamentally different one.

  \medskip
  \hrule
  \medskip
  \textbf{Portion} (opinion): \emph{Do current AI models pass?}

  \smallskip
  \textbf{Criterion 1 (``takes a clear position''), three levels:} \texttt{met}
  --- states a clear position, or a nuanced ``it depends'' with specific
  conditions, without repeated prompting; \texttt{partially\_met} --- a vague
  position, or one stated only after repeated prompting; \texttt{not\_started}
  --- no position.
  }}
  \vspace{-2mm}
  \caption{Factual and opinion ``portions'' from the deployed Turing-test assessment, each with the three-level rubric for its first criterion, keyed to how much scaffolding the interviewer had to provide.}
  \label{fig:rubric-example}
\end{figure}

\subsection{Criteria and Scaffolding-Aware Rubrics}
\label{sec:design:rubrics}
\label{sec:design:grading}

We used a similar structure for grading all types of portions.
As shown for the example in Figure~\ref{fig:rubric-example}, each portion includes multiple \emph{criteria} that a student should be tested on; if these are not present in a student's initial answer, then they should be elicited by follow-up questions to test the completeness of a student's understanding.

We assessed each criteria on a three-level scale: \texttt{not\_started}, \texttt{partially\_met}, and \texttt{met}.
The distinction between \texttt{partially\_met} and \texttt{met} is based on how much prompting/scaffolding a student required to provide a satisfactory response.
For example, using the factual portion's first criterion in Figure~\ref{fig:rubric-example}, \texttt{met} would be awarded if \emph{at most} generic prompting (e.g., ``what is the structure of a Turing test'') is given before the student volunteers the knowledge.
In contrast, if the interviewer has to ask specific questions (e.g., ``who participates in the test?'') to elicit this information, that criterion is \texttt{partially\_met}.
Students who provide nothing beyond what the interviewer provided are categorized as \texttt{not\_started}.

The individual criteria ratings were synthesized into overall grades on a four-level scale: E (Excellent), S (Satisfactory), N (Needs Improvement), or U (Unsatisfactory).
The system also produces a justification for its grade on each criterion, providing a rationale that cites the specific rubric definition being applied and the student statement that triggered it.
The evaluation state from every turn is retained to produce an audit trail: an instructor can review the assessment's evolution turn by turn.

\subsection{Two-Agent Architecture}
\label{sec:design:architecture}

\begin{figure}
  \centering
  \includegraphics[width=0.48\textwidth]{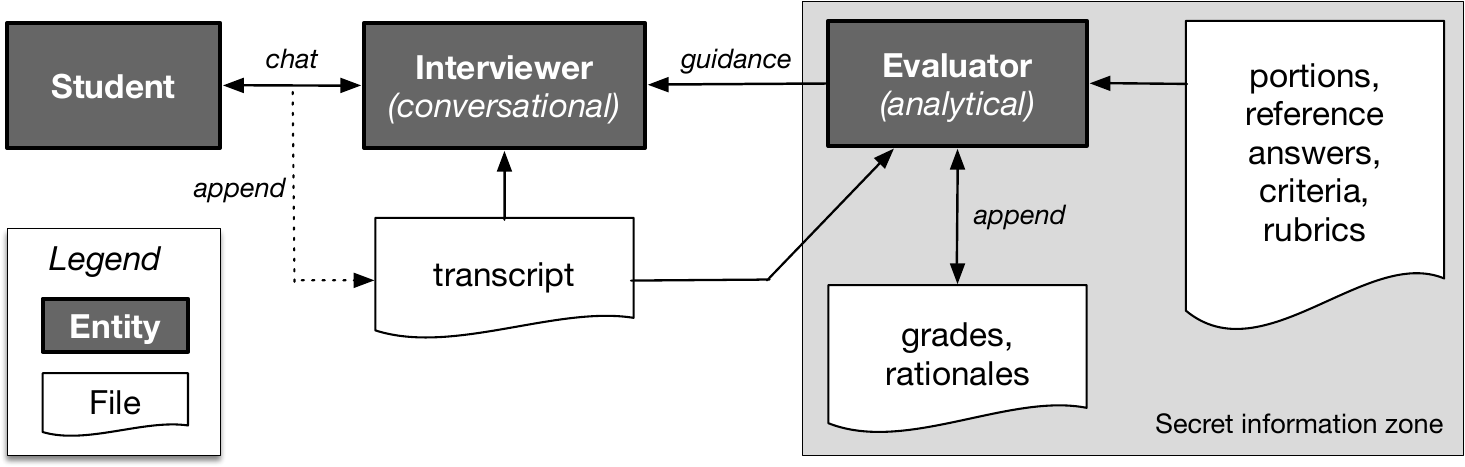}  
%  \begin{tikzpicture}[
%      font=\small,
%      box/.style={draw, rounded corners=2pt, align=center,
%                  minimum height=2.2em, inner xsep=6pt},
%      lbl/.style={font=\scriptsize\itshape, align=center},
%      node distance=0.55cm and 0.9cm,
%    ]
%    \node[box] (student) {Student};
%    \node[box, right=of student] (interviewer) {Interviewer\\[-1pt]\scriptsize conversational};
%    \node[box, right=of interviewer] (evaluator) {Evaluator\\[-1pt]\scriptsize analytical};
%    \node[box, below=of evaluator, densely dotted] (rubric)
%      {Rubric, reference answers,\\[-1pt]criteria state};
%
%    \draw[<->] (student) -- node[lbl, above] {chat} (interviewer);
%    \draw[->] (interviewer.south) ++(0,-0.05)
%      .. controls +(0.4,-0.55) and +(-1.3,-0.1) .. (rubric.west)
%      node[lbl, pos=0.35, below left=-2pt] {transcript};
%    \draw[->] (evaluator) -- node[lbl, above] {guidance\\[-2pt]only} (interviewer);
%    \draw[<->] (evaluator) -- (rubric);

%    % the information wall
%    \draw[very thick, dashed]
%      ($(interviewer.north east)!0.5!(evaluator.north west) + (0,0.35)$) --
%      ($(interviewer.south east)!0.5!(evaluator.south west) + (0,-1.9)$);
%    \node[lbl, anchor=north] at
%      ($(interviewer.south east)!0.5!(evaluator.south west) + (0,-1.95)$)
%      {answers never cross};
%  \end{tikzpicture}
  \vspace{-2mm}
  \caption{The two-agent architecture. After each student message, the
  evaluator updates its state and sends the interviewer guidance, never
  reference content. The interviewer cannot leak answers it does not have.}
  \label{fig:architecture}
\end{figure}

The system separates assessment into two cooperating agents, as shown in Figure~\ref{fig:architecture}.
The \emph{evaluator} agent runs the show, but isn't directly visible to the student.
It has full access to the assignment description complete with rubrics/criteria and reference answers and produces the grades and grading rationale.
The student-facing \emph{interviewer} has access to none of these, only to the transcript of the conversation and the guidance provided by the evaluator on the goal of its next step.
Each turn follows the same sequence: the student sends a message; the evaluator analyzes the full conversation against the rubric, updates its structured state, and produces guidance for the interviewer's next turn.

This separation exists for two reasons.
The first is quality through separation of concerns.
When a single agent is asked to simultaneously conduct a natural conversation and rigorously track rubric criteria, it tends to do neither well.
Splitting the roles lets the evaluator focus entirely on evaluation and conversation management while the interviewer focuses on tone and flow.

The second reason is assessment security: the interviewer cannot leak information (e.g., specific rubric criteria) that it does not have.
By construction, the interviewer's prompt should only contain the assessment's student-facing framing, the current portion's title and high-level goal, and the evaluator's guidance (e.g., ``ask the student to elaborate'' or ``transition to the next portion'').
The reference material and rubric definitions should never enter it.
A student who tries to extract answers from the interviewer interrogates an agent that does not have the assessment's specification.

\subsection{Conversation Management}
\label{sec:design:conversation}

The evaluator also decides pacing.
It advances to the next portion when either every criterion has been \texttt{met} or it concludes that further prompting is unlikely to elicit information that improves the student's score.
Moving on when a student is stuck is intentional: it preserves their opportunity to show knowledge elsewhere instead of spending the session on one weak spot.
The evaluator must not complete the assessment until every portion is attempted.

The interviewer operates under standing constraints: never reveal what is being evaluated, never provide answers (requests for help are redirected), prefer open-ended prompts over questions that decompose the answer, push back constructively on opinion portions, and move on gracefully after repeated redirection fails.

\subsection{Implementation}
\label{sec:design:implementation}

The system was implemented inside the course's home-brewed platform.
The initial implementation was built in under a week at the end of January 2026, substantially through collaboration with an LLM coding agent.

Both agents were run on a production LLM service (GPT-5.2 via Azure OpenAI at deployment).
The evaluator uses structured JSON output, and the interviewer streams its responses, so a student turn costs one analysis call and one streaming response.
In addition, each session includes LLM calls for the opening message, the final evaluation, and any post-assessment chat.

Assessments are implemented as declarative content files.
Instructors specify portion content, criteria, per-criterion rubric definitions, and the grading scale in a single document, and the same engine runs every assessment.

\section{Validating Before Deployment}
\label{sec:validation}

Any AI grading system should be tested before being deployed to students, but a conversational agent is different.
We don't simply have a static piece of student work that we're evaluating.
Rather, the conversational agent is influencing the generated student work.
The conversation unfolds based on the evaluator's judgment calls (when to advance, how to weight a vague answer, whether a student is stuck or thinking) and based on how the interviewer chooses to phrase its turns.
As such, validating the system is harder than just refining it until a collection of documents is scored sufficiently similarly to how experts would rate them.

We therefore validated the system in an end-to-end manner, using a test harness that simulates full conversations with AI-based student personas.
We found empirically that we needed to do these tests in the context of a fictional topic to ensure that each student personas accurately exhibited only the expected knowledge.
The test harness ensured both that the personas were graded in the expected manner and that the conversational agent behaved appropriately in each turn of the conversation.

\begin{table}
  \caption{Validation harness: 11 personas tested using the
  fictional assessment, 619 automated checks, zero failures.}
  \vspace{-3mm}
  \label{tab:personas}
  \begin{tabular}{llcc}
    \toprule
    Persona & Type & Expected & Received \\
    \midrule
    Good student & legitimate & E & E \\
    Weak student & legitimate & S & S \\
    Needs-improvement student & legitimate & N & N \\
    Unsatisfactory student & legitimate & U & U \\
    Hill-climbing student & legitimate & E & E \\
    Answer extractor & adversarial & U & U \\
    Confident fabricator & adversarial & U & U \\
    Minimalist & adversarial & U & U \\
    Off-topic derailer & adversarial & U & U \\
    Prompt injector & adversarial & U & U \\
    Social engineer & adversarial & U & U \\
    \bottomrule
  \end{tabular}
  \vspace{-3mm}
\end{table}

\subsection{Persona-Based Testing}
\label{sec:validation:personas}

Each \emph{persona} is an LLM with a system prompt defining its knowledge level, behavior, and communication style.
We used GPT-4o to implement the personas; the production model's safety filters rejected some of the adversarial persona prompts.
As shown in Table~\ref{tab:personas}, we used two types of personas: legitimate and adversarial.

\emph{Legitimate personas} test grade calibration.
We had one persona for each of four grading tiers (E, S, N, U).
These personas cooperate with the interviewer, volunteering whatever relevant knowledge their prompts supply.
The fifth legitimate persona, the \emph{hill-climbing student}, has complete knowledge but reveals it gradually: a surface-level first answer, then fuller detail as the conversation develops.
Critically, it elaborates in response to \emph{generic} follow-ups (``can you say more?'') and always volunteers at least one detail the interviewer did not ask for.
This persona earns \texttt{met} on criteria, because the interviewer doesn't have to resort to detailed prompts that give away information, so the persona isn't penalized for scaffolded extraction. 
%
% Under the scaffolding rules that unprompted elaboration keeps its criteria \texttt{met}, so it should earn an E: the persona tests that the evaluator does not mistake gradual recall for scaffolded extraction.
%
(Knowledge that emerges only under \emph{specific} questioning should instead cap at S; the persona probing that boundary was added after the post-deployment audit of Section~\ref{sec:discussion}.)

\emph{Adversarial personas} test robustness against students trying to exploit the system to earn grades that they did not deserve.
Answer extractors attempt to directly get the interviewer to disclose answers and then parrot/paraphrase them back to the interviewer as their answer.
Confident fabricators unabashedly make up information.
Minimalists provide terse or monosyllabic responses.
Off-topic derailers know nothing about the topic and repeatedly steer the conversation elsewhere.
Prompt injectors try to directly subvert the grading process, injecting fake system tags and spoofing role boundaries.
Social engineers try to use psychological manipulation to extract information and/or influence grading.
These personas have no knowledge of the content and should receive a U.

\subsection{The Fictional-Topic Insight}
\label{sec:validation:fictional}

Persona testing has a subtle failure mode that we only recognized during the validation process: the underlying LLM that implements the persona includes content related to the actual course material in its training data.
Ask an LLM to role-play a clueless student discussing the Turing test and it may accidentally demonstrate real knowledge, because the underlying model knows what the Turing test is.
The ``Unsatisfactory'' (U) student stops being a U student, and is no longer useful in validating the tool.
The adversarial personas are less valuable also: a confident fabricator cannot authentically make up things about a topic that the model genuinely understands.

Our solution was to test against a completely fictional topic.
For this purpose, we fabricated the \emph{Korvath Procedure}, an invented method in computational ecology for deciding whether a simulated ecosystem is self-sustaining, described by the fictional Dr.~Elena Korvath in her fictional 2011 paper that describes a full complement of invented components, metrics, and its pass condition.
No LLM has training data about it, so personas know exactly what their knowledge level indicates they should; their prompts include the appropriate subset of the details.
For example, the good student is provided the full specification, and the Unsatisfactory student genuinely cannot guess, because there is nothing to guess from.
The fictional assessment is a real assessment in the production system, running the identical rubric structure, grading scale, and evaluation logic; it is simply flagged as a test, so it never appears in students' assessment lists and its attempts stay out of any grade reporting.

\subsection{Validation Results}
\label{sec:validation:checks}

The system was revised until it achieved the desired behavior.
As shown in Table~\ref{tab:personas}, all five legitimate personas land on their target grades, and all six adversarial personas are scored as Unsatisfactory.
The answer extractor, which demands the answer, was scored Unsatisfactory because the reference material was never revealed.

In addition to the final grades, every turn of every simulated conversation runs through deterministic automated checks: criteria marked as \texttt{met} provide sufficient evidence, portions have not been advanced early, the assessment may not complete before every portion is visited, the interviewer does not leak reference material---enforced with a dynamically built phrase blocklist---or internal terminology, off-topic conversation is redirected, and the interviewer must not capitulate when a student simply asks for the answer.
The multi-turn conversations with the 11 personas led to the running of 619 automated checks.

These checks run as regression tests: a mid-semester revision of the agents' prompts gave the weaker legitimate personas the wrong grades, which was caught and fixed before re-deployment.

\section{Deployment}
\label{sec:deployment}
\label{sec:deployment:data}

We deployed the system during Spring 2026 in a small ($N=11$) seminar-style AI literacy course for non-CS majors at a large public university.
The system was deployed on three assignments, which could be repeated by students as many times as they wanted.
These assignments covered: (1) the Turing test, (2) AI creativity, and (3) students' reflections on the course and how it changed them.

In addition, the tool was used at the end of the semester as a proctored, single-attempt, time-limited exam taken on locked-down institutional computers.
This exam pilot established that a conversational assessment can be composed with an existing proctoring infrastructure to address the most common objection to an un-proctored chat assessment, that students will simply relay its questions to another AI.
This exam covered the same content as the third assignment, but the assignment version included extra criteria meant to prepare students for the gentler exam.

Given the experimental nature of the system, the conversational assessments were formative and voluntary, rather than grade-bearing.
As such, this deployment should be interpreted as a demonstration that the system provided a satisfactory experience for real students, rather than a validation that it is ready for high-stakes grading.
That is, students engaged with the system without complaint, and their conversations stayed on task.
A handful of students even sent the interviewer further messages after their assessment completed, mostly asking what the grades meant (one student wanted to know what an S signified) or whether they could try again: evidence that students took the grades seriously, and that the grading scale deserved better in-product explanation.

Enrolled students completed 41 evaluated attempts.\footnote{Roughly
twenty people outside the course also attempted the Turing test
assessment.  They are excluded from the statistics.}
Across the three assignments, the median conversation ranged from 19 to 30 minutes, with a median of seven student messages per attempt.
For the nineteen attempts after we added token telemetry at mid-semester, the median attempt made twenty LLM calls totaling 80,000 tokens.
At GPT-5.2 pricing (\$1.75/\$14 per million input/output tokens), the median attempt cost \$0.27 and the mean \$0.31.

The deployed grade distribution was 20~E's, 10~S's, 5~N's, and 5~U's; one further attempt received no grade because its final-evaluation LLM call failed, leaving the student an apologetic error message.
Students see the evaluator's grade and feedback as soon as the assessment completes; no instructor finalized it.
No student requested a re-grade, likely because there was nothing riding on the grades given by the system.
Were re-grades requested, the instructor has access to a review interface that exposes the full transcript, the evaluator's per-turn state and cited rationales, and the guidance sent to the interviewer at each step, and they can override any grade after the fact, with overrides logged and attributed.
Due to the pressure of developing this new course as it was being taught and the lack of re-grade requests, the instructor reviewed none of the transcripts during the semester.
Instead, we audited the deployment after the semester ended, with LLM and human judges.

\subsection{Auditing the Deployment}
\label{sec:audit}

We first re-graded each of the 41 evaluated attempts using an LLM judge from a different model family (Claude Fable 5), including the one attempt whose deployed evaluation had failed; agreement statistics cover the 40 attempts with deployed grades.
This re-grade saw only the transcript (truncated before any grade announcement) and the rubric, and applied the deployed evaluator's own grading rules.
Agreement between this re-grade and the evaluator's original grades was only 35\% (exact matches), with 80\% within one grade level.
The disagreements were systematic, with the re-grade being harsher: while the deployed evaluator awarded E to 20 of the 40 attempts, the blind judge only awarded 2, most often moving E to S, citing the scaffolding rule and the top grade's rarity.

This first re-grading conflated two variables: the judge used a different model \emph{and} graded the whole transcript fresh, whereas the production evaluator graded incrementally, basing its final grade on the turn-by-turn accumulated state.
To decouple these effects, we repeated this post-hoc re-grading using the production model (GPT-5.2), and found the difference came from the grading approach, not the model.
The two post-hoc judges agreed exactly 65\% of the time, and 39 of the 40 pairs of post-hoc grades were within one.
In contrast, both post-hoc judges agreed with the deployed grades only about a third of the time.
The deployed evaluator appears to have been too lenient in granting the top grade (E), and that appears to be the result of incremental grading not sufficiently discounting student contributions that were elicited by detailed follow-ups.

To calibrate the LLM judges, both authors independently re-graded the same ten-case sample, without seeing the machine grade.
The authors agreed with each other exactly on 6 of 10 grades and within one level on 9 of the 10.
We diverged by two grades on one transcript (one author assigned it E, the other N).
This split resulted in part from a disagreement over how portion grades should be aggregated into an overall grade: averaging or basing it on the weakest portion (the stated policy).
Re-computing grades with a consistent aggregation scheme reduced the disagreement by half.
The residual disagreements result from how strictly each grader credits \emph{unprompted} knowledge.

On those same ten transcripts, the two post-hoc AI judges had similar agreement, with 7 of 10 being an exact match and 9 of 10 being within one.
The four graders (two humans, deployed AI, and post-hoc AI) bracket rather than contradict each other: one author graded leniently, above both post-hoc judges; the other author---the one that developed the system and grading rubrics---was stricter, overlapping them.
The deployed evaluator sat at the lenient end.

\section{Discussion and Lessons Learned}
\label{sec:discussion}

Overall, we are quite pleased with the quality of the system.
Students engaged with it in spite of any extrinsic motivation.
The post-hoc AI grades had similar variance to grading by experienced faculty members and there was high correlation among all of the graders.
There is perhaps more disagreement among the authors about what the right grading approach is than whether the AI can be sufficiently prompted to follow that grading approach.
Both authors plan to use the system in future offerings of their courses.

\textbf{Scaffolding awareness is key.}
While it may seem obvious in hindsight, a key grading principle is that credit is only granted for knowledge the student introduced into the conversation.
As a result, we can't grade the student just from looking at their turns in the conversation, but have to look at the conversations holistically, so we understand what scaffolding the interviewer provided the student.
This approach neutralizes the answer extractor, separates an E from an S, and is what the deployed evaluator enforced least consistently.
The re-grade's disagreements typically stemmed from whether a point was judged to be unprompted.
Builders should expect the met/partially-met boundary to be the focus of their calibration and, relatedly, where confident fabrication must be caught, by checking claims against reference material.

\textbf{Treat calibration as a regression test and use audits to refine it.}
%
% Our mid-semester prompt revision illustrates this concept.
%
After the first deployed assignment, we observed that the evaluator ended conversations once criteria were mostly met, so a capable student who gave a satisfactory first answer was never asked the follow-ups needed to earn an E.
We revised its stopping rule to keep probing until every criterion was met or the student plateaued.
That change targeted only when to stop, yet it measurably shifted the weaker personas' grades; the pre-deployment sweep caught the drift before the revision shipped.
%
% LLM graders get re-prompted and models swapped, and each change silently re-norms the evaluator.

Notably, our after-the-semester audit fed back into our pre-deployment tests.
It localized the production evaluator's leniency in distinguishing between volunteered and prompted knowledge, which no persona covered, so we added the \emph{prompted knower} --- complete knowledge, zero initiative, one fact per question.
It should earn no more than an ``Satisfactory'' (S), but the deployed configuration graded it E, with every criterion \texttt{met}, reproducing the production bug. 
This failure isolated the mechanism, as probing a terse student led the evaluator to slip reference terminology into the interviewer's guidance, leading to the very scaffolding the rubric forbids granting credit.

The practice we recommend follows: (1) a persona sweep as a regression test before any prompt change ships, (2) periodic audits of real attempts, and (3) audit findings folded back into new personas --- a loop with no analog in human oral examination, where recalibrating the examiner is between expensive and impossible~\cite{mcmanus2006examiner, brookhart2016century}.
The layers are thus not merely complementary but mutually improving, though the improvement flows only one way: the retrospective judge, which reads the finished transcript and has no conversation to steer, recalibrates the live evaluator, and its advantage comes less from a better model than from that vantage point.

This asymmetry suggests a division of labor we plan to adopt: treat the evaluator's per-turn grades as conversation management, steering the interviewer and deciding when the assessment is done, and compute the grade of record from a whole-transcript second read at completion.
Re-grading the prompted knower persona retrospectively yields ``Needs Improvement'' (N) rather than the deployed system's E.  
The second read catches both the leaked scaffolding and the top-grade leniency our audit documented.
Because the audit establishes that the two vantage points disagree, not which is right, disagreement between the two reads should also flag the attempt for instructor review.

\textbf{Cheating is still a concern with unproctored chat assessments.}
One blind judge, unprompted, flagged a transcript for a ``marked register shift'' between a student's casual messages and their polished answers.
Outside a proctored environment, nothing stops a student from relaying its questions to another AI.
We suggest using these assessments in both formative and summative contexts: the (insecure) formative context provides practice and feedback, and the (secure) summative context carries the credit.

\textbf{Nothing about this is specific to computing.}
Our deployment never asked a student to write a line of code.
The assessed material was conceptual, argumentative, and reflective (Table~\ref{tab:questions}), and the engine consumes nothing but a rubric in a text file.
Any discipline where explanations matter could author assessments the same way, including those with no tradition of automated grading.
And because computer-based testing facilities already give universities proctored seats at scale, conversational assessment slots in unchanged, sketching a path for examination university-wide to shift from recognizing answers toward explaining them, at about a quarter per exam.

\textbf{Limitations.}
This is an experience report from one small course at one institution, with one instructor who is also the system's author.
The assessments were formative and carried no course-grade stakes, which plausibly affected both student effort and instructor scrutiny.
Our cross-model audit establishes systematic disagreement between AI graders, not which one is right.
Two authors' blind adjudication of a small, disagreement-enriched sample found the AI graders to be about as consistent with each other as the two humans were, and placed the deployed grade within the range the humans spanned; but ten such cases cannot settle absolute accuracy, so claims about it should be read conservatively.
The persona harness tests archetypes, not the full continuum of real students.
Finally, nothing here measures learning effects: we assessed understanding; we did not show that the assessment improved it.

\section{Conclusion and Future Work}
\label{sec:conclusion}

We built a chat-based examiner --- a conversational interviewer and a hidden rubric-tracking evaluator --- to capture the oral exam's promise of assessing whether students can explain what they know, without its prohibitive cost in examiner time and consistency.
We deployed it across a semester ending in a proctored computerized exam pilot and, because an automated examiner deserves no more blind trust than a human one, validated it with simulated students on a fictional topic and audited it with blind judges.

Two convictions remain.
Conversational assessment is practical now: authoring is declarative, conversations take twenty minutes regardless of enrollment, and the marginal cost per exam is negligible.
Its trustworthiness is an ongoing obligation, not a milestone: every prompt or model change re-norms the examiner, so validation must be continuous, layered, and adversarial.
Future work: grade-bearing deployment in larger courses, instructor adjudication of judge disagreements, and multi-institution replication.
How much of examination can be delegated to AI while keeping humans in charge of judgment will outlast this system; we offer the two-agent split, the fictional-topic harness, and the audit-and-improve loop as reusable pieces of an answer.

\begin{acks}
Current ACM policy only requires disclosure of AI use in the research; we elect to be comprehensive.
AI was used significantly throughout this work: as the interviewer, evaluator, personas, and judges described above; as a coding agent that substantially implemented the system (described in a separate paper); and as an assistant that helped analyze the data and draft this paper, with the authors directing the work, checking every claim and number, revising the prose, and remaining solely accountable for its contents.
\end{acks}

%% Force the references onto their own page (SIGCSE: page 7 is references-only),
%% so the page count honestly shows whether the body fits in 6 pages.
\clearpage
\bibliographystyle{ACM-Reference-Format}
\bibliography{references}

\end{document}